\documentclass[11pt,twocolumn,oneside]{article}

\newif\ifpdf
\ifx\pdfoutput\undefined
\pdffalse 
\else
\pdfoutput=1 
\pdftrue
\fi

\ifpdf
\usepackage[pdftex]{graphicx}
\else
\usepackage{graphicx}
\fi

\renewcommand{\section}[2]{}
\newcommand{\heada}[1]{\stepcounter{section}\begin{center}{\bf \arabic{section}. #1}\end{center}}
\newcommand{\headb}[1]{{\bf #1}}

\title{{\bf Wikis in Tuple Spaces}}
\author{\bf G Gordon Worley III\\\bf School of Computer Science, University of Central Florida\\\bf Orlando, FL 32816 USA}
\date{}

\begin{document}

\ifpdf
\DeclareGraphicsExtensions{.pdf, .jpg, .tif}
\else
\DeclareGraphicsExtensions{.eps, .jpg}
\fi

\maketitle

\thispagestyle{empty}

\begin{center}{\bf ABSTRACT}\end{center}

We consider storing the pages of a wiki in a tuple space and the effects this might have on the wiki experience.  In particular, wiki pages are stored in tuples with a few identifying values such as title, author, revision date, content, etc. and pages are retrieved by sending the tuple space templates, such as one that gives the title but nothing else, leaving the tuple space to resolve to a single tuple.  We use a tuple space wiki to avoid deadlocks, infinite loops, and wasted efforts when page edit contention arises and examine how a tuple space wiki changes the wiki experience.


\heada{INTRODUCTION}

A wiki is an easily editable hypertext system.  It's like a Web site, except that any visitor to a wiki page can also edit the page.  Ward Cunningham created the first wiki in 1995.  Since then, wikis have gradually grown in popularity.  Today hundreds of wiki engines exist, each putting its own spin on the general wiki idea.\cite{Wikipedia05}  The most popular wiki on the Web, Wikipedia, receives an average of 11 million hits a day.\cite{Wikimedia04}

All wikis store data as pages with unique names.  These names are then used to reference the pages.  Sometimes these pages are stored in databases, other times as individual files, but always there is one active version of a page and a linearly ordered history of zero or more previous revisions.\cite{Wikipedia05}  While this fits with familiar document access metaphors, it can lead to problems because wikis are expected to handle not just concurrent viewing, like most Web pages, but also concurrent editing, where possibly two or more people are trying to edit the same page at the same time.

Most wiki engines resolve these conflicts by always accepting each editor's changes, even if it overwrites an unseen revision.  Some wiki engines warn editors that they're about to overwrite unincorporated changes, while others try to merge the editors' revisions, relying on a human to do the final work of stitching the two together.\cite{Cunningham05}  In wikis with many users, though, these tasks can cause problems as the number of users who might try to edit the same page grows.  Eventually, a wiki could lock up with multiple editors trying to modify the same page, annihilating each other's changes or trapped in a loop of merging, submitting, and merging.

Assuming we don't want to eliminate concurrent editing, we consider the use of tuple spaces to store wiki pages to improve the situation.  Tuple spaces are concurrent systems invented by David Gelernter that have a bag of k-tuples, each capable of holding arbitrary values.  Tuple spaces are accessed by the functions OUT, IN, and RD.  OUT adds a tuple into the tuple space, IN removes a tuple by matching a pattern against the tuples in the tuple space, and RD works like IN but returns a copy of a tuple without removing it from the tuple space.  Tuple spaces get their properties by accepting all OUTs and resolving INs and RDs to return any one tuple that matches the template rather than a particular tuple.\cite{Carriero89}

We show a way to implement a wiki using tuple spaces, rather than databases or flat files.  First we consider how to reimplement editing and reading, then consider the effects of a tuple space wiki on the wiki experience, and finally look at some extensions to and open questions about tuple space wikis.

\heada{IMPLEMENTATION}

In a tuple space wiki, pages become collections of similar tuples.  Reading is a matter of searching for a tuple with certain properties, and editing is a matter of putting a tuple into the tuple space.  We first consider the editing process and then the reading process.

\headb{Editing}\\Tuple spaces provide two ways to edit.  One is to IN a tuple from the tuple space, modify it, and then OUT it to the tuple space.  The other is to RD a tuple, modify it, and OUT it to the tuple space, so both the original and the modified version reside in the tuple space.  The former editing style is undesirable, because it doesn't allow concurrent editing and could result in a reader fault---a reader wants to find a certain page and can't because it's currently out of the tuple space.  Aside from the annoyance to readers, this could also lead to duplicate pages as readers choose to create pages that they think don't exist but actually do.  Therefore, we consider only the latter editing style.

Let's step through the editing process.  First Ed, an editor, accesses the wiki by searching the tuple space for a page (this process is described in detail in the next subsection).  If his search doesn't return a tuple, and assuming the search was well formed, he's discovered a page that doesn't exist.  He may choose to create this page by constructing a wiki tuple and OUTing it to the tuple space.  Since OUT always succeeds, if another editor, say Alice, notices the missing page and constructs a similar tuple at the same time, the two OUTs will not conflict, and the tuple space will contain two tuples representing the page.

Now consider that Ed's search does return something.  He reads the page and decides to modify it some.  He makes his modifications to the tuple and OUTs it to the tuple space.  Again, because of the properties of OUT, Alice can construct a similar tuple at the same time without contention, and Ed's version of the tuple will join the page's existing tuples.

As is probably becoming clear, what we mean by page in a tuple space wiki is different from what is normally meant.  A tuple space wiki `page' is really a set of tuples, tied together by some similarities.  Editing a page always expands the page's set of tuples.  How we read pages, and how tuples are bound into pages, is defined by template-match searches.

\headb{Reading}\\Robert, a reader, wants to access a particular page in the wiki.  To do so, he RDs the tuple space and gets back a tuple if his RD succeeds.  RD takes a template of the tuple he wants as an argument.  The tuple space takes this template and finds a tuple that matches it without regard for how many tuples may match the template or which tuple of all the tuples that match the template is returned.  Let's look closer at what this means.  Consider that every tuple in our tuple space wiki has the following form:

{\tt (wikiword, author, rev number, date modified, body)}

And that we have the following tuple space:

{\tt \{(TupleSpace, Ed, 1, 2005-03-20, Tuples are great!), (TupleSpace, Alice, 2, 2005-03-22, Tuples are indeed great.), (Ed, Alice, 1, 2005-03-20, Ed is my friend.)\}}

If Robert wants to access the TupleSpace page, he might call {\tt RD((TupleSpace, ?, ?, ?, ?))}, where question marks are unspecified values in the template.  The tuple space will go out and return either {\tt (TupleSpace, Ed, 1, 2005-03-20, Tuples are great!)} or {\tt (TupleSpace, Alice, 2, 2005-03-22, Tuples are indeed great.)}, but there's no way to know ahead of time which one Robert will get, or even that both exist without directly examining the tuple space.  From Robert's perspective, he issues an RD and gets back a tuple and nothing else.

Robert could also issue more exact searches, such as {\tt RD((TupleSpace, Alice, ?, ?, ?))}, which returns a tuple written by Alice and titled TupleSpace, or different forms of searches, like {\tt RD((?, Alice, ?, ?, ?))}, which returns a tuple written by Alice.  The defined characteristics of a template provide a way identifying groups of tuples we can call pages.  Traditionally, pages are identified by what in our example is the wikiword variable in the first position in the tuple, and we consider all the tuples with the same wikiword to make up a page, but we also get author pages, date pages, etc. based on each piece of metadata stored in the tuple that we can search with a template.  Templates redefine pages as a bag of instances rather than a linear order of instances.

\begin{figure}
\begin{center}
\begin{tabular}{c}
\resizebox{3 in}{!}{\includegraphics{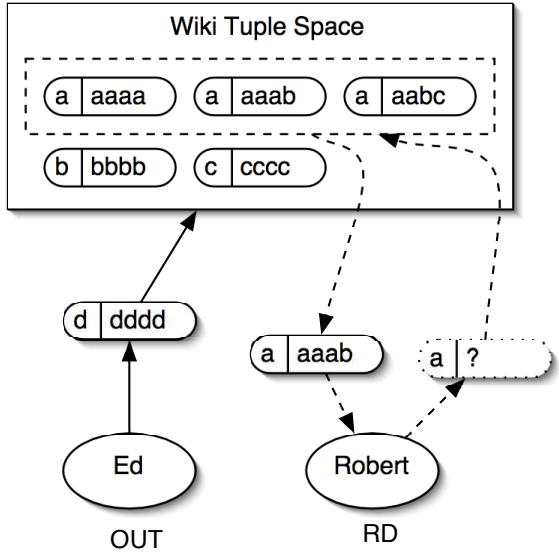}}
\end{tabular}
\end{center}
\caption[example] 
{ \label{wtsfig}
\small Another example of a tuple space wiki.  Here Robert performs an RD by sending the tuple space a template, to which the tuple space returns a copy of a matching tuple among all possible matches.  Ed OUTs a new tuple to the tuple space.
}
\end{figure}

Thus we now have a simple way to redefine wiki page storage using tuple spaces.  An additional, graphical example can be found in Fig. \ref{wtsfig}.  Now we turn to effects on the wiki experience and extensions to our ideas.

\pagebreak
\heada{EFFECTS AND EXTENTIONS}

As described, a tuple space wiki allows an unlimited number of editors to concurrently edit the same page without contention.  As wikis grow in size, this may prove very important.  Consider Wikipedia, which already has thousands of editors and millions of readers.  Given that even comparatively small wikis see edit contention \cite{Cunningham05}, Wikipedia likely already suffers from edit contention problems on a small scale.  As the number of readers grows into the hundreds of millions and billions and the number of editors grows into the millions, edit contention will become a serious issue that, left unaddressed, could cripple Wikipedia's progress.  The same will apply to any other wiki that grows to large size, or has a high ratio of editors to pages.

Aside from better handling of concurrent editing, a tuple space wiki is arguably more democratic than a linear wiki, where the pages are a sequence of linearly stored revisions.  Although linear wikis already provide a very democratic form of communication \cite{Masum04, Cedergren03}, readers of a linear wiki usually only see the latest revision of each page, though many previous revisions may exist, but in a tuple space wiki every previous revision has a chance of being seen by the nature of the RD operation.  Such fairness to previous revisions, while preventing editors from snuffing out opinions and alternative views, could annoy readers with excessively frequent low quality tuples, such as ones containing typos or vandalism, that have already been edited and improved.  It seems tuple space wikis need a way to return only the best tuples for a particular page to a reader without eliminating valuable content.

One way to approximate a return-best RD is by changing the probability that a particular tuple that matches a template will be returned.  Inspired by agoric algorithms \cite{Wildavsky04}, and without needing to modify tuple spaces, we may be able to achieve such a system in the following way.  Each time someone reads a page, they are given the option to vote for the tuple they saw.  Voting for a tuple simply means OUTing a new copy of it to the tuple space, thus increasing the number of instances of that tuple by one.  On the next RD of the same page, the voted for tuple will be more likely to return than it was before.  Eventually, one or several good tuples for that page should become popular enough that they will be more likely to appear than all other tuples in that page.  As the number of voted copies of tuples in a page increases, the chances of getting an unpopular tuple decrease towards zero.

But this approximation is not perfect.  It may be possible for a person to game this system, repeatedly OUTing a tuple of it until it becomes the most likely to return when reading a page, not because it was the most popular tuple, but because it was the most voted for.  Gaming could be used to vandalized a page or suppress valid content, and because users can only vote for, not against, tuples, this may not be easily reversible without administrators stepping in to purge the tuple space.  A better system might allow unvoting, but this wouldn't eliminate gaming, just provide a mechanism to fight it, and the fighting might lead to violent swings in the popularity of particular tuples as opposing parties battle to vote and unvote a tuple.  Such chaos is undesirable to most readers and editors, who want to work on content, not play politics.

Alternatively, readers might try to coax a tuple space wiki page into a linear structure by searching for pages with added date information.  This may work well for recently edited pages, but putting date information in the template may make infrequently edited pages appear not to exist.  To counteract this, readers might set the date farther and farther back, but the earlier the date, the more tuples that might match, until every tuple matches as if the date were ignored.  This is similar to the way linear wikis obtain a best revision, but it doesn't take full advantage of the tuple space and still relies on time to find the best tuple, which isn't necessarily a good indicator of quality.

This is just a first look at effects of and extensions to tuple space wikis.  There's much more to be explored.
\pagebreak
\heada{CONCLUSIONS}

Tuple space wikis provide an alternative means of storing pages, producing a different wiki experience for users, where edit contention is not an issue and pages are distributions of tuples sharing some criteria rather than linearly ordered revisions.  Although this may lead to greater democratic sharing of ideas, readers may not be willing to tolerate the misinformation and vandalism that would be more viewable in a tuple space wiki.

There are not yet any wiki engines, and thus no wikis, that use tuple spaces.  An open line of research is to implement such a wiki engine, run a wiki on it, and see how users react.  This may lead to tweaking of the tuple space wiki idea, keeping the benefits while finding ways to reduce the negatives.  Another open problem is to develop a good return-best RD, which will probably require experimentation in real systems to find a solution that works with the user community.

\begin{center}{\bf REFERENCES}\end{center}

\bibliographystyle{unsrt}
\bibliography{wikits}

 \end{document}
 \end